\documentclass{mem}
\usepackage{natbib}
\usepackage{amssymb}
\usepackage{txfonts}
\usepackage{balance}
\usepackage{graphicx}
\usepackage[a4paper]{hyperref}
\idline{75}{282}
\begin{document}

\title{
Reconstructing the evolution of dark energy with variations of fundamental parameters
}

   \subtitle{}

\author{
N.\,J.~Nunes\inst{1}, 
T.~Dent\inst{2}, 
C.\,J.\,A.\,P.~Martins\inst{3}
\and
G.~Robbers\inst{4}
          }


\institute{
Institut f\"{u}r Theoretische Physik, Philosophenweg 16, 69120 Heidelberg, Germany  \\
\email{nunes@thphys.uni-heidelberg.de}
\and
School of Physics and Astronomy, Cardiff University, The Parade, Cardiff CF24 3AA, U.\,K. 
\and
Centro de Astrof\'{\i}sica, Universidade do Porto, Rua das Estrelas, 4150-762 Porto, Portugal, and
DAMTP, University of Cambridge, Wilberforce Road, Cambridge CB3 0WA, U.\,K.
\and
Max-Planck-Institut f\"ur Astrophysik,
Karl-Schwarzschild-Stra{\ss}e 1, D-85748 Garching bei M\"unchen, Germany
}

\authorrunning{Nunes et al.}
\titlerunning{Reconstructing dark energy}

\abstract{
Under the assumption that the variations of parameters of nature and the current acceleration of the universe are related and governed by the evolution of a single scalar field, we show how information can be obtained on the nature of dark energy from observational detection of (or constraints on) cosmological variations of the fine structure constant and the proton-to-electron mass ratio. We also comment on the current observational status, and on the prospects for improvements with future spectrographs such as ESPRESSO and CODEX.
\keywords{Cosmology: dark energy, quintessence, variation of alpha, variation of mu}
}
\maketitle{}

\section{Introduction}
We propose to probe the dynamics of the equation of state of matter and energy in the universe by using observations of cosmologically varying fundamental parameters. This requires a number of assumptions: first, that there is clear evidence for dark energy that could be attributed to a rolling scalar field or quintessence; second, that there is a cosmological variation of the fine structure constant of magnitude $\Delta\alpha/\alpha \lesssim 10^{-5}$, as suggested by Keck/HIRES high resolution quasar absorption spectra (\cite{Murphy:2003mi,Murphy:2003hw}, but see also \cite{Srianand:2007qk}); and thirdly, that any variation of $\alpha$ arises from the evolution of the quintessence field (\cite{Dvali:2001dd,Chiba:2001er}). 

Concerning the first assumption, there is currently little observational evidence for dark energy to be more than a bare cosmological constant. However, if dark energy indeed results from an evolving scalar field, it could be expected to couple to other forms of matter and lead to variations of masses and couplings (implying the second and third assumptions) unless some unknown symmetry principle explicitly forbids these couplings.

We take the coupling between the scalar field and electromagnetism to be 
${\cal L}_{\phi F} = - \frac{1}{4} B_F(\phi) F_{\mu\nu}F^{\mu\nu}$ where the gauge kinetic function $B_F(\phi)$ is linear, 
$ B_F(\phi) = 1- \zeta \kappa (\phi-\phi_0)$ (and $\kappa^2=8\pi G$). This can be seen as the first term of a Taylor expansion, and should be a good approximation if the field is slowly varying at low redshift. Then, the evolution of alpha is given by
\begin{equation}
\frac{\Delta \alpha}{\alpha} \equiv \frac{\alpha-\alpha_0}{\alpha_0} =
\zeta \kappa (\phi-\phi_0) \,.
\end{equation}

We can also consider the variation of $\mu\equiv m_p/m_e$. In grand unified theories we expect a correlation between the variation of $\alpha$ and $\mu$ (\cite{Calmet:2001nu}) given by $\Delta \mu/\mu = R\, \Delta \alpha/\alpha$, where $R$ is a model-dependent numerical factor arising from correlated variations of $\Lambda_{QCD}$, the Yukawa couplings, the vacuum expectation value of the Higgs field and $\alpha$ itself. Under simple assumptions we obtain $R \sim -20$, which is in severe tension with observations that indicate a nontrivial variation of $\alpha$ at high redshift but null variation of $\mu$ (\cite{King:2008ud,Thompson}) with equal or better precision. This simple exercise illustrates the potential of cosmological observations of quasar absorption lines and variation of fundamental parameters in discriminating particle physics models.

\section{Reconstruction procedure}
In order to test our third assumption we may verify whether a given model of dark energy proposed in the literature can fit the data. This has been done by a number of authors (\cite{Copeland:2003cv,Anchordoqui:2003ij,Dent:2008vd,Bento:2008cn}) and it was found that many models can satisfy all the constraints, though only in islands of the parameter space. 

An alternative approach is to extract the quintessence scalar potential from the observational evolution of the data on the variation of $\alpha$. In \cite{Parkinson:2003kf}, it is assumed that both the form of the gauge kinetic function $B_F(\phi)$ and of the equation of state parameter $w_\phi(z)$ are known. The authors parametrize these functions and fit the parameters by combining quasar and SnIa data.

Here we discuss a slightly different approach (\cite{Nunes:2003ff,Avelino:2006gc,Avelino:2008dc}). We parametrize 
$B_F(\phi)$ with a linear function as above and parametrize the evolution of $\alpha$ and/or $\mu$ with a polynomial $g(N)$ where $N = - \ln(1+z)$. Essentially all we need is a functional form of $\phi'(N)$, then we integrate the equation
\begin{equation}
\sigma' =  -(\kappa\phi')^2 (\sigma + a^{-3}) \,.
\end{equation}
where $\sigma = \rho_{\phi}/\rho_0\Omega_{M0}$. The solution $\sigma(N)$ then provides the evolution of the equation of state parameter through
\begin{equation}
w  = -1 + \frac{(\kappa \phi')^2}{3} \left( 1+
\frac{1}{\sigma a^3} \right)  \,,
\end{equation}
There are thus three steps of the reconstruction procedure required before we can apply these equations.

\subsection{Step 1: Obtaining the data sets}
The first step consists of obtaining data on the evolution of $\alpha$ and/or $\mu$, either from existing observations, or from simulated data, for the purpose of forecasting the accuracy of reconstructions with proposed future spectrographs. We will consider the second approach. The data are generated from the numerical evolution of the quintessence field for a specific scalar potential $V(\phi)$. We consider the normal distribution with mean $\Delta \alpha/\alpha = \zeta \kappa (\phi-\phi_0)$ where $\Delta\alpha/\alpha(z = 3) = -0.5 \times 10^{-5}$ and we chose $R = -6$. We have assumed that with the {\tt ESPRESSO} spectrograph for VLT, 200 systems will be found to determine $\alpha$ and 50 objects to determine $\mu$ with sensitivity $\delta = 5 \times 10^{-7}$. With the {\tt CODEX} spectrograph for the E-ELT, we consider 500 systems measuring $\alpha$ and 100 constraining $\mu$ with sensitivity $\delta = 10^{-8}$.

\subsection{Step 2: Fitting the data}
In previous works we chose to fit the data with a polynomial
$g(N) \equiv \Delta \alpha/\alpha  = g_1 N + g_2 N^2 + ... + g_m N^m$, then the velocity of the field is simply $\kappa \phi' = g'/\zeta$.

\subsection{Step 3: Estimating $\zeta$}
\begin{figure}
\includegraphics[width=3.2cm]{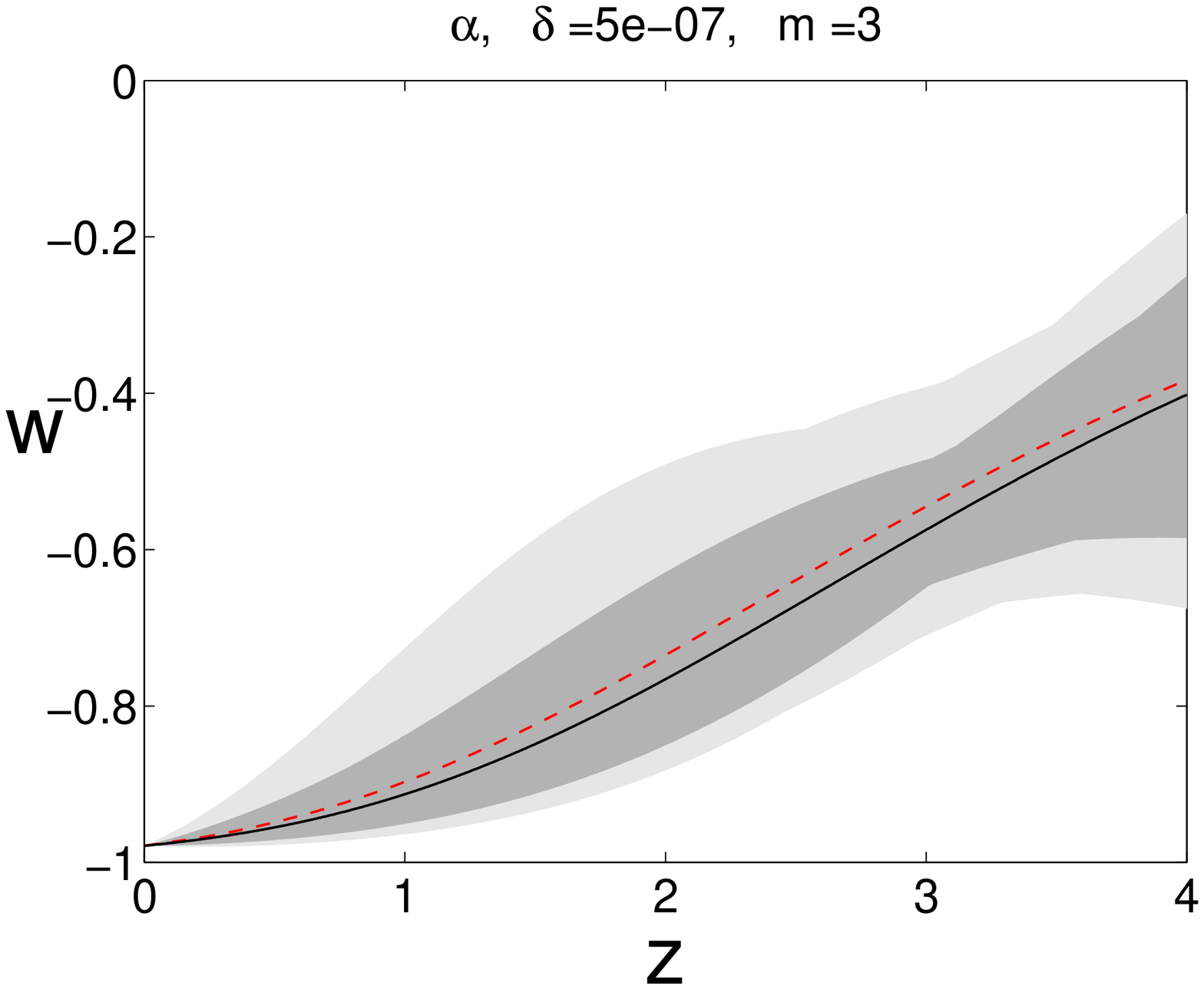}
\includegraphics[width=3.2cm]{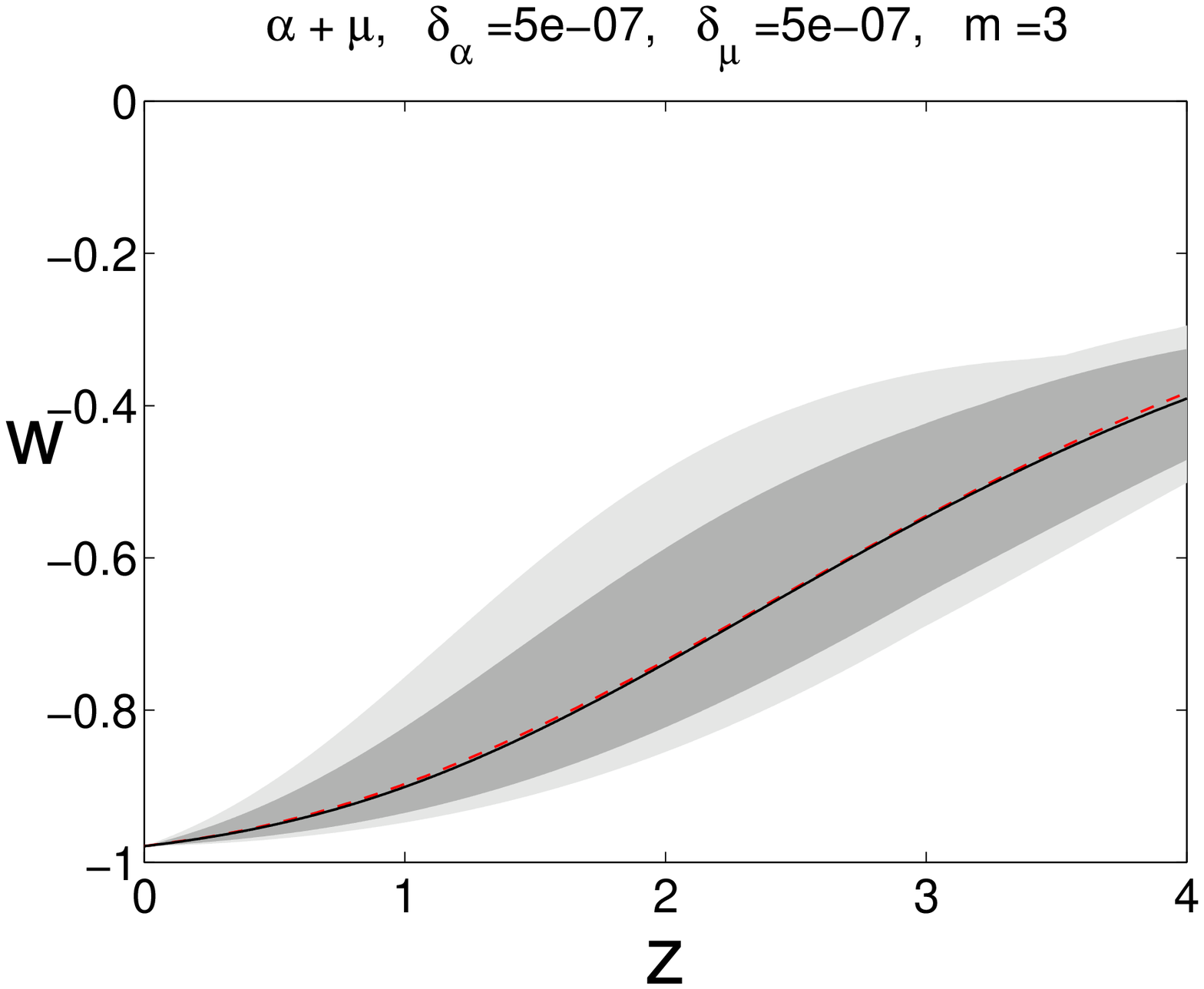} \\
\includegraphics[width=3.2cm]{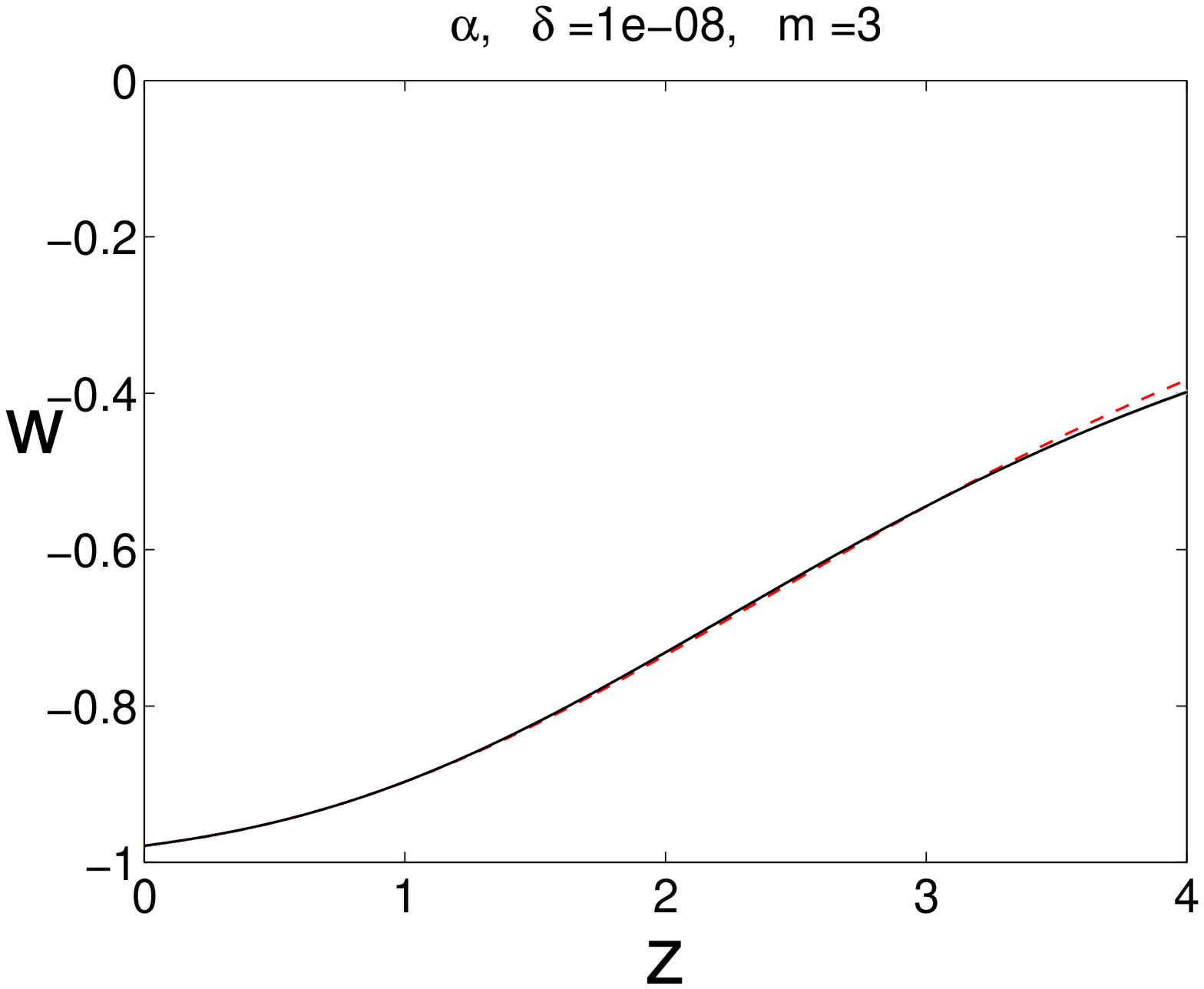}
\includegraphics[width=3.2cm]{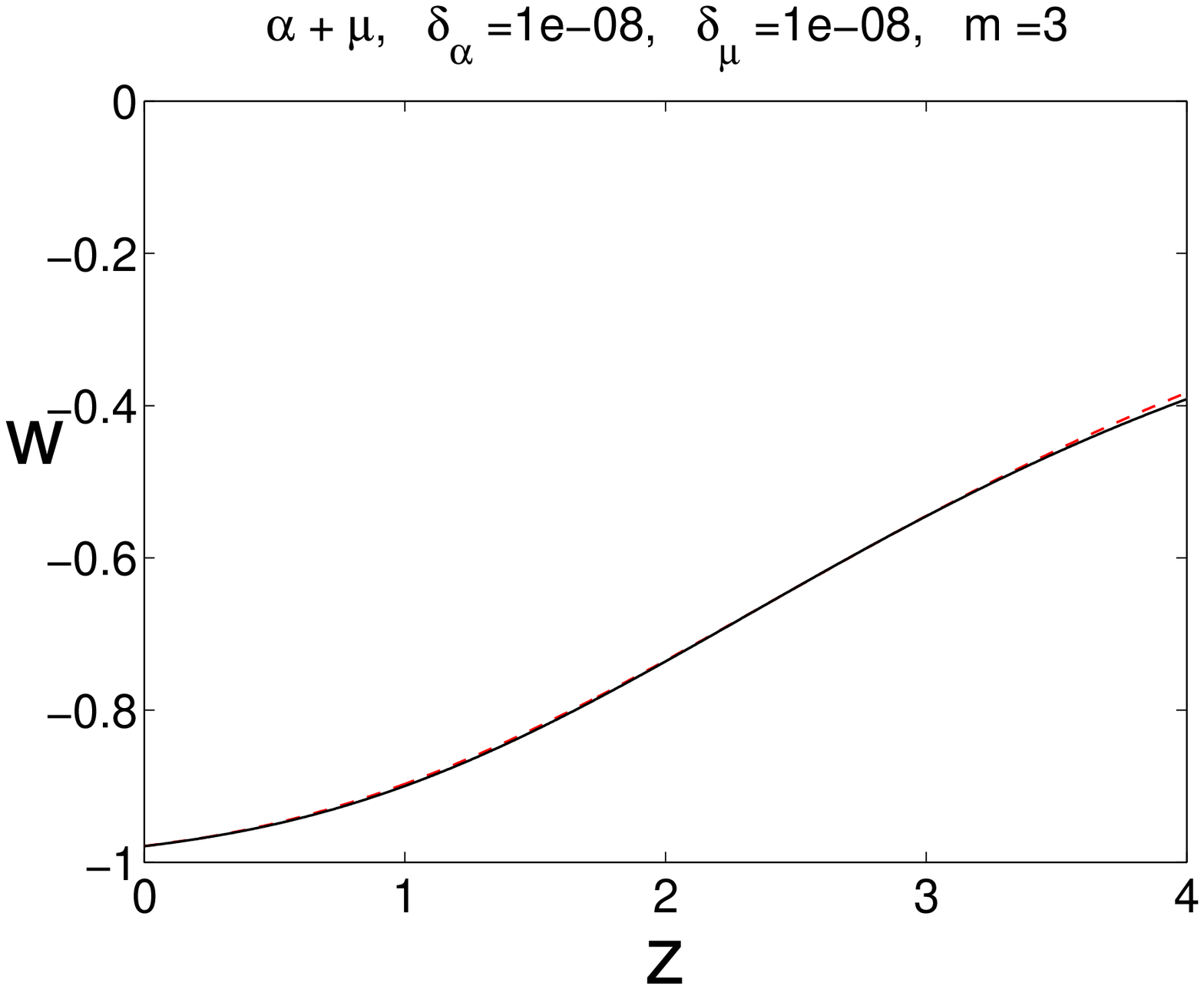}
\caption{\label{reconstrucoes} Reconstruction results for the scalar potential $V(\phi) = V_0 (e^{10 \kappa \phi} + e^{0.1 \kappa \phi})$. The dashed line represents the fiducial model, the solid line the best fit reconstruction and the dark and light bands the corresponding 1$\sigma$ and 2$\sigma$ errors. Upper left panel: using $\alpha$ measurements with {\tt ESPRESSO}; upper right panel: using $\alpha$ and $\mu$ with {\tt ESPRESSO}; lower panels illustrate the same reconstructions with {\tt CODEX}.}
\end{figure}

The only missing ingredient is the value of $\zeta$. We must estimate its value from independent observations such as SnIa or weak lensing. 
More specifically, from the relation
\begin{equation}
w = -1 + \frac{(\kappa\phi')^2}{3 \Omega_\phi} \,,
\end{equation}
and substituting for $\kappa \phi'$ in terms of $g'(N)$ we can obtain $\zeta$ at any redshift. For example, at redshift $z = 0$ we have 
\begin{equation}
\zeta^2 = \frac{1}{3} \, \frac{g_1^2}{\Omega_{\phi 0} (1+w_0)} \,.
\end{equation}
For typical values of ${\Omega_\phi}_0 \approx 0.7$,  $w_0 \sim [-0.99, -0.6]$ and $g_1 \sim 10^{-5}$ we obtain $\zeta \sim 10^{-7} - 10^{-4}$ which is comparable to bounds resulting from tests of the weak equivalence principle (\cite{Olive:2001vz,Dent:2008vd}).
In Fig.~\ref{reconstrucoes} we illustrate a reconstruction example for {\tt ESPRESSO} and {\tt CODEX} using $\alpha$ alone, and $\alpha$ and $\mu$ data in combination. 

\section{The Rosenband bound}
A strong constraint on the current variation of $\alpha$ was obtained recently using atomic clocks (\cite{2008Sci}) 
\begin{equation}
\dot{\alpha}/\alpha = (-1.6 \pm 2.3) \times 10^{-17} {\rm yr}^{-1} \,.
\end{equation}
This result rules out many models of quintessence with a monotonic evolution of the field with a linear coupling if $\Delta\alpha/\alpha(z = 3) \sim 10^{-5}$, including the example of the previous section. There are of course a number of ways of evading these bounds. For instance, one may consider an oscillating evolution: a potential with a minimum, $V = V_0(\exp(10 \kappa \phi) + \exp(-0.5 \kappa \phi))$ would now satisfy the Rosenband bound because there is an oscillation of the field when this reaches the minimum of the potential. 

Alternatively one can modify the gauge kinetic function (\cite{Marra:2005yt}). For example if we consider the following gauge kinetic function
\begin{equation}
B_F = 1- {\zeta} (\phi - \phi_0)^{q} \,,
\end{equation}
then our procedure would still apply but now there are two parameters which must be determined using independent data. The following relations may be used:
\begin{eqnarray}
\frac{1}{q} &=& 1- \frac{g \,g''}{g'^2}  \\
&~& + \frac{3}{2} \frac{g}{g'} \left( \frac{w'}{3(1+w)} + w (\Omega_\phi-1)\right) \,, \\
\zeta^{2/q} &=& \frac{1}{q^2} \left(\frac{g'}{g}\right)^2 \frac{g^{2/q}}{3\Omega_\phi(w+1)} \,.
\end{eqnarray}
These require knowledge of the slope $w'(z)$ and second derivatives of the polynomial, $g''(z)$ at a given redshift. The reconstruction is therefore less accurate and specially difficult if $w_0 \approx -1$.

\section{What do the current data tell us?}
In this section we are going to be even more open minded by trying to dismiss our theoretical prejudices and simply seek to take the current data at face value and understand what they might be telling us.

Let us take the unbinned \cite{Murphy:2003hw} data which suggest a smaller value of $\alpha$ in the past. Considering the Rosenband bound and in addition the Oklo and meteorite analysis which put bounds of $\Delta \alpha/\alpha = (0.7 \pm 1.8) \times 10^{-8}$ at redshifts $z=0.14$ (\cite{Gould:2007au}) and $\Delta\alpha/\alpha = (1.5 \pm 1.5)\times 10^{-6}$ (\cite{Olive:2003sq,Dent:2008gx}) at $z = 0.45$, we may be compelled to consider a sharp transition in the value of $\Delta \alpha/\alpha$ at about redshift $z =1$ (see also \cite{Mortonson}). With this in mind we propose to keep the linear dependence of the gauge kinetic function $B_F(\phi)$ but introduce the following parametrization for the evolution of the scalar field
\begin{equation}
\label{parameq1}
\phi - \phi_0 = c \left[ \tanh\left(\frac{N-N_t}{\Delta}\right) - 
                  \tanh\left( -\frac{N_t}{\Delta}\right)\right] \,.
\end{equation}
\begin{figure}
\includegraphics[width=4.8cm]{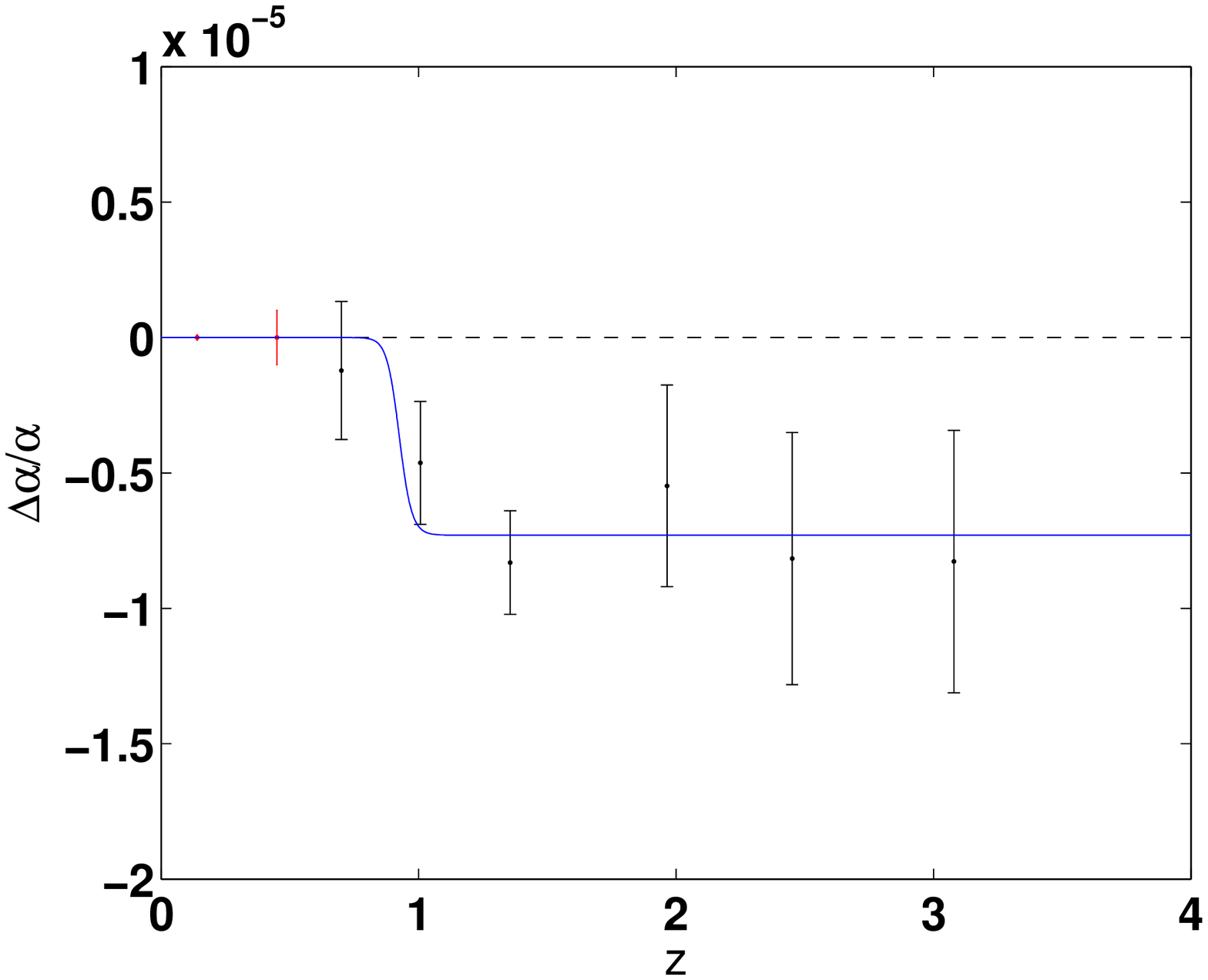} \\
\includegraphics[width=3.2cm]{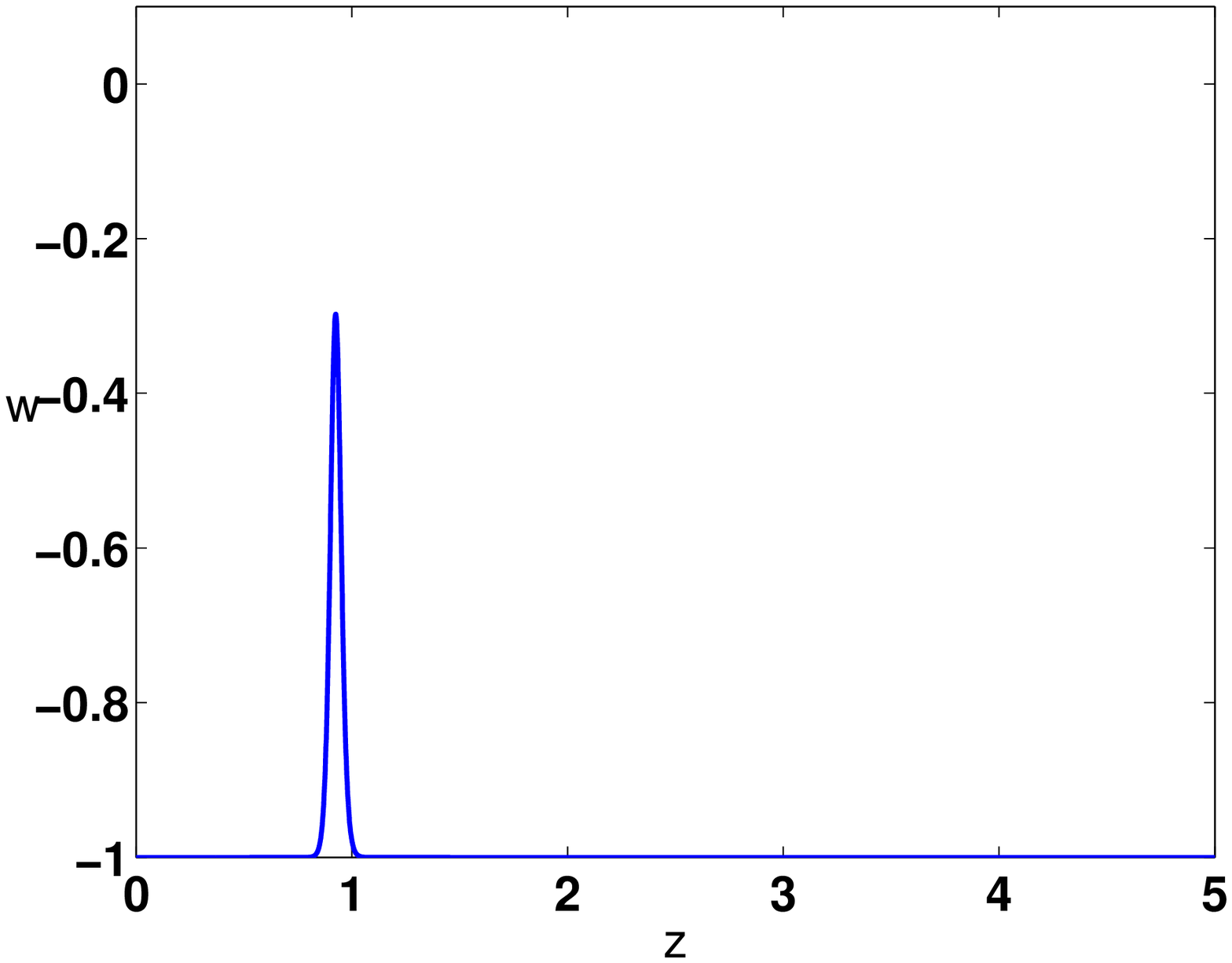} 
\includegraphics[width=3.2cm]{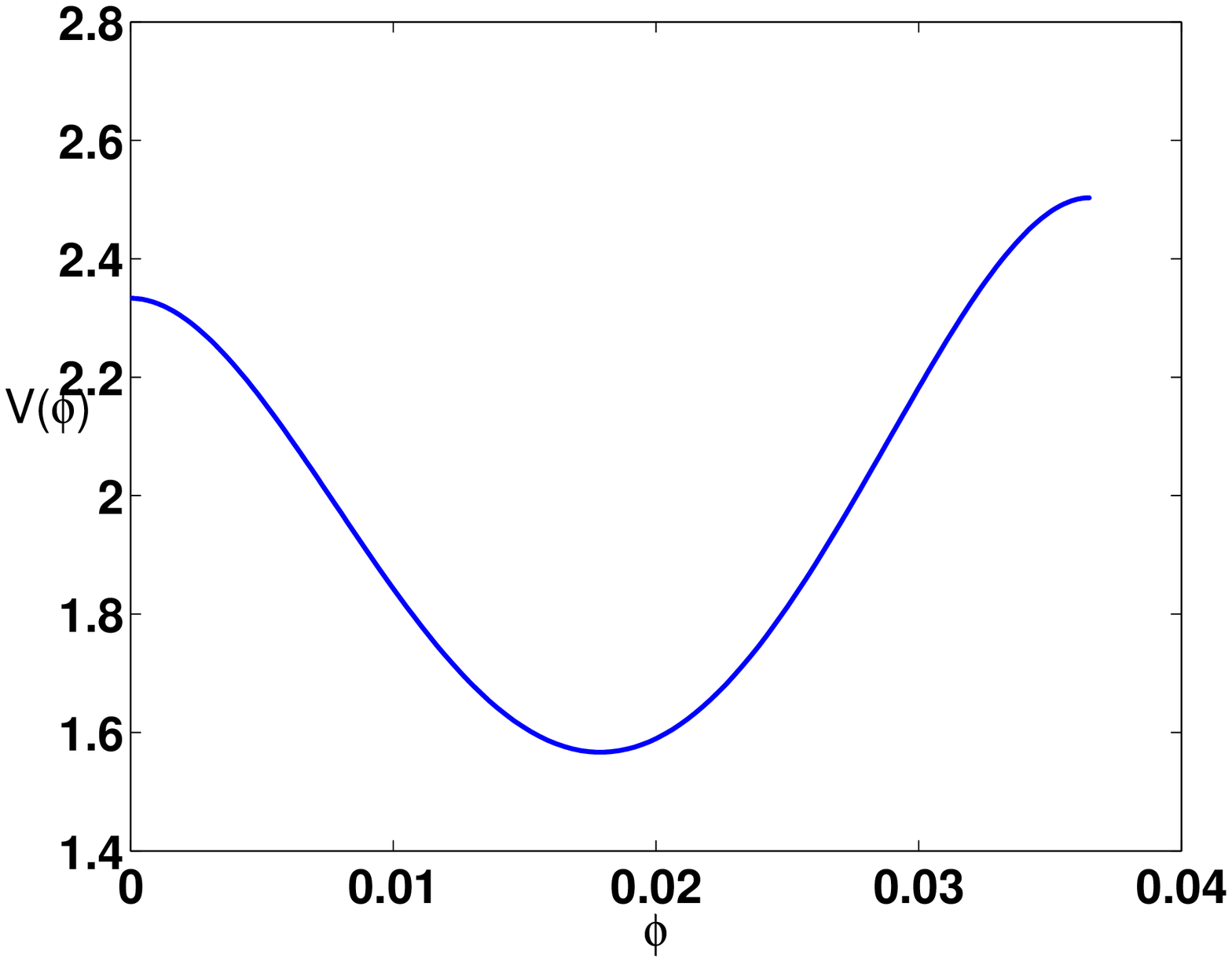}
\caption{\label{param1} Parametrization (\ref{parameq1}). Upper panel: comparison with binned data; lower left: evolution of the equation of state parameter $w(z)$; lower right: the scalar potential.}
\end{figure}
This parametrization corresponds to a field that evolves from a local maximum of the scalar potential, falls in a steep well and rises again approaching another local maximum (see Fig.~\ref{param1}). The velocity of the field is therefore decreasing today and a large vacuum energy is attained.

An alternative parametrization is a function that allows for a variation of the field at early times
\begin{equation}
\label{parameq2}
\phi - \phi_0 = c \,\frac{N}{N_t} \left[ \tanh\left(\frac{N-N_t}{\Delta}\right) - 
                  \tanh\left( -\frac{N_t}{\Delta}\right)\right] \,. 
\end{equation}
\begin{figure}
\includegraphics[width=4.8cm]{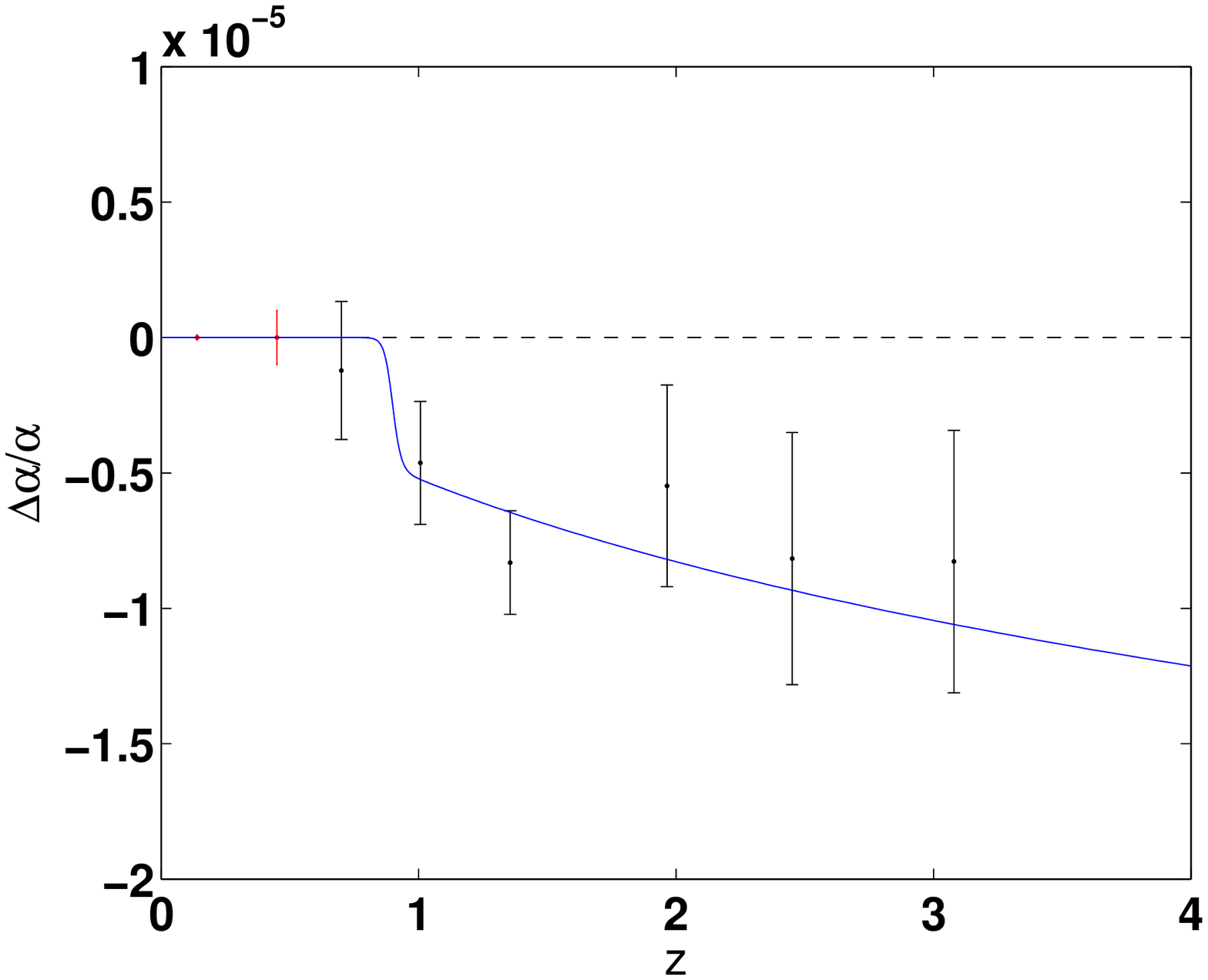} \\
\includegraphics[width=3.2cm]{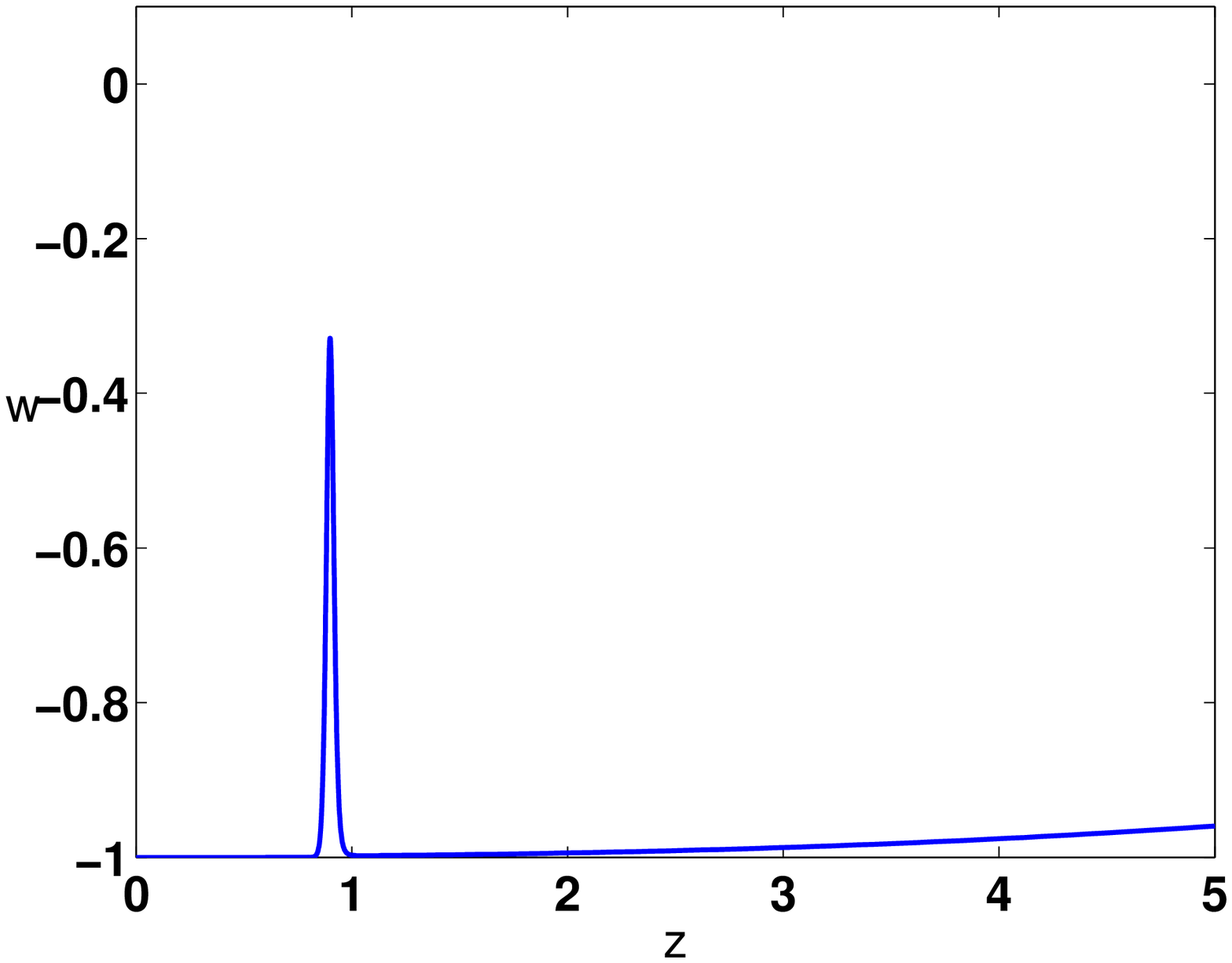} 
\includegraphics[width=3.2cm]{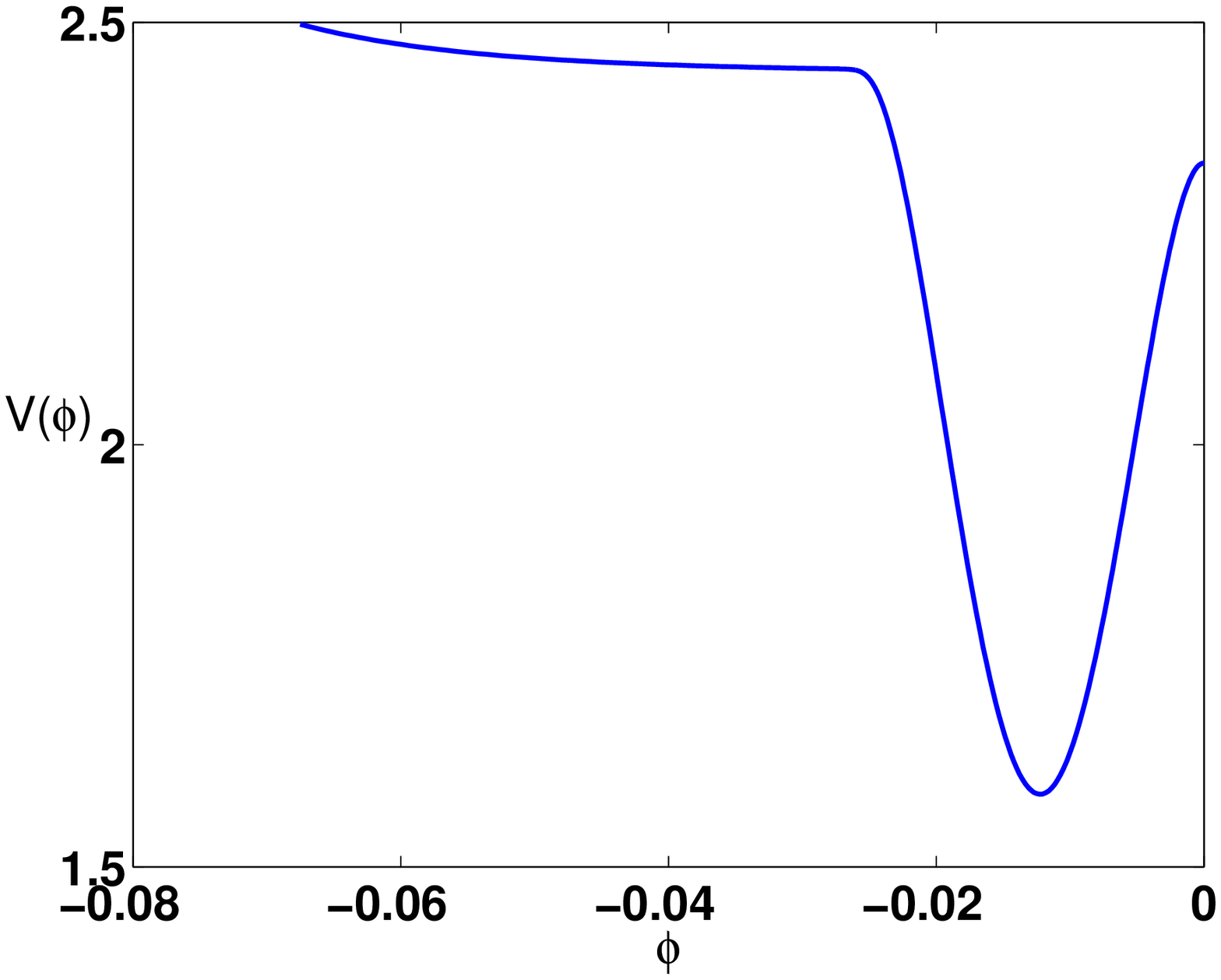}
\caption{\label{param2} Parametrization (\ref{parameq2}). Upper panel: comparison with binned data; lower left : evolution of the equation of state parameter $w(z)$; lower right panel: the scalar potential.}
\end{figure}
This second parametrization does not require that the field is initially at a local maximum, but instead allows an equation of state parameter that approaches $w(z) \approx  0$ at large redshifts (see Fig.~\ref{param2}). These forms of the potential look distinctly unnatural, however we emphasise that our objective here is to disregard theoretical prejudices and to use the data almost blindly in order to uncover viable forms of the scalar potential. On the other had, this simple exercise highlights the importance of an independent observational confirmation of these variations.

\begin{figure}
\includegraphics[angle=-90,width=6cm]{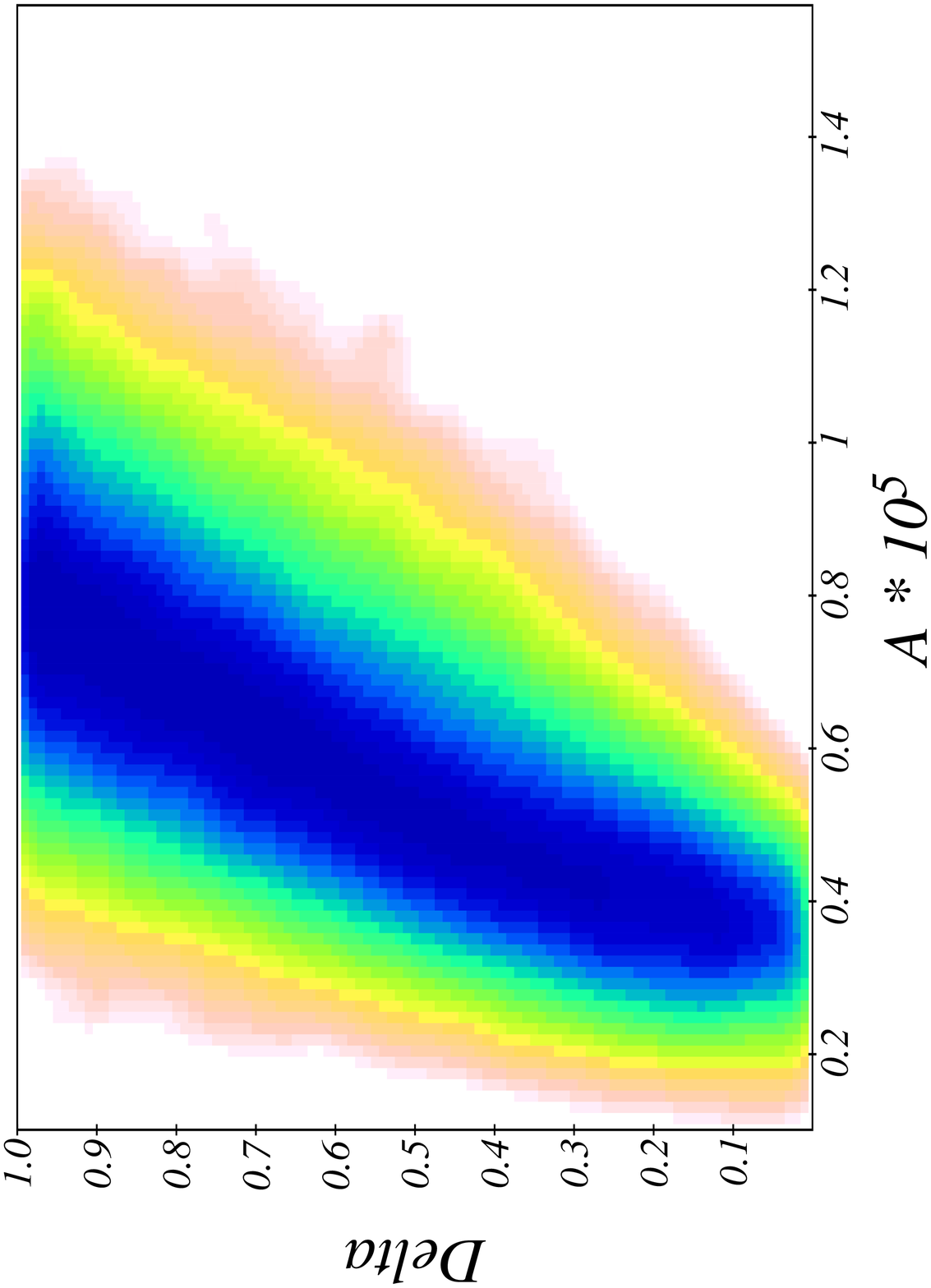} \\
\includegraphics[angle =-90,width=6cm]{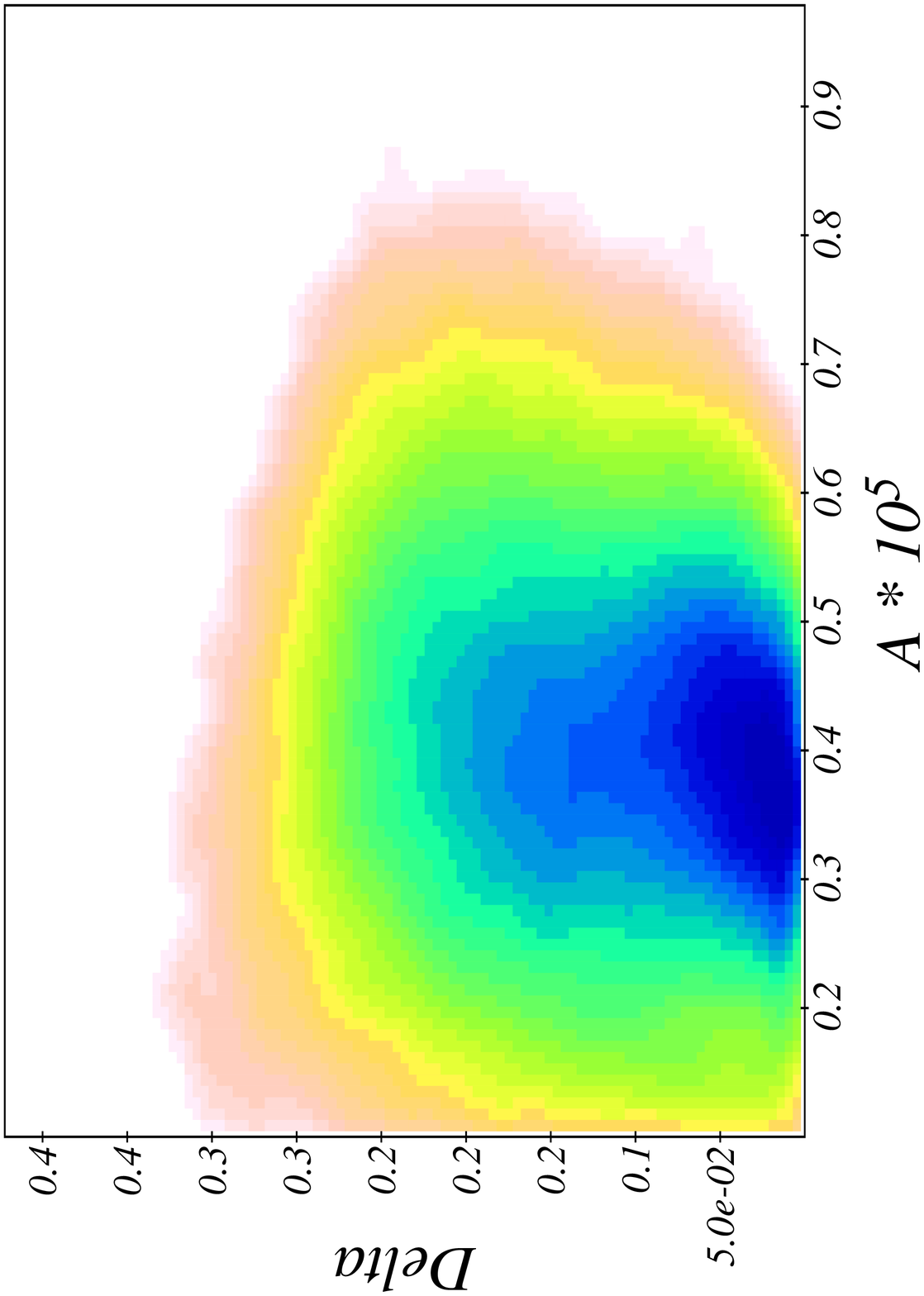}
\caption{\label{contornos} Likelihood analysis for parameters $A$ and $\Delta$ using only quasar data (upper panel) and quasar, Oklo and meteorites data and the Rosenband bound (lower panel). The several shaded regions represent 1$\sigma$, 2$\sigma$ and 3$\sigma$ confidence regions.}
\end{figure}

Performing a likelihood analysis using the first parametrization we obtain the contour plots shown in Fig.~\ref{contornos}, for the amplitude of the transition in $\Delta \alpha/\alpha$, $A = c \zeta$ and the width of the transition, $\Delta$. We observe that by including extra constraints such as the Rosenband bound, Oklo and meteorites, the contours are tighter.

\begin{figure}
\includegraphics[width=6cm]{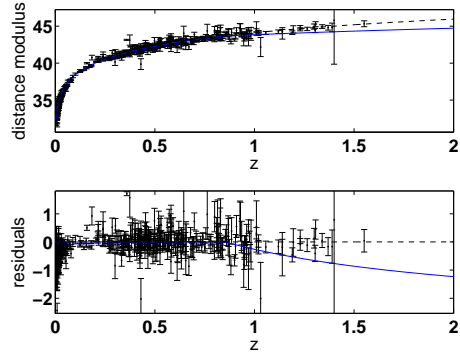} 
\caption{\label{supernova} The effect of different values of $\zeta$ on the luminosity distance for a fixed value of $A =c\zeta$. The dashed line represents a quintessence model with large $\zeta$ which is completely degenerate with a $\Lambda$CDM cosmology and the solid line represents a quintessence model with small $\zeta$. We used here the parametrization in Eq.~(\ref{parameq1}).}
\end{figure}

Combining the data and constraints on the variation of $\alpha$ with SnIa data we are then able to constrain $c$ and therefore $\zeta$. Indeed, to large values of $\zeta$ correspond small values of $c$ and therefore the luminosity distance is to all effects indistinguishable form a $\Lambda$CDM scenario. The only upper limit on $\zeta$ comes from tests of the equivalence principle. Small values of $\zeta$, however, give an evolution deviating substantially from $\Lambda$CDM and one should be able to put lower bounds on this quantity with cosmological data at redshift $z > 1$, as illustrated in Fig.~\ref{supernova}.

\section{Conclusions}
We have shown that under simple assumptions we can determine the nature of dark energy, not by {\it fitting}\/ the parameters of a scalar potential to cosmological data, but by performing the inverse procedure, that consists in using quasar data to {\it reconstruct}\/ the potential. We have seen that the evolution of the equation of state parameter can in principle be found, subject of course to the precision of future data. These observations have profound implications as the simple knowledge of the sign of $w'(z)$ can help us to favour or discard freezing models of quintessence (increasing $w(z)$ with increasing redshift), thawing models (decreasing $w(z)$) and k-essence models (typically also decreasing $w(z)$).

This type of reconstruction directly probes the scalar field dynamics and may be carried out, with current data, to redshifts beyond $z = 4$, far higher than the limiting value $z = 1.7$ of SnIa searches. Future astrophysical techniques may extend this to even higher redshifts (\cite{Levshakov,Kozlov}). Moreover, the observations can be done from the ground and are consequently cheaper than satellite-based observations.

Here, we have presented the reconstruction procedure for a minimally coupled scalar field, but other models with non-canonical kinetic terms, couplings to matter or multiple fields might have further interesting phenomenological properties and therefore lead to alternative approaches.

\begin{acknowledgements}
NJN is supported by Deutsche Forschungsgemeinschaft, project TRR33 and thanks the organisers of the IAU and the JD9 discussion section for a stimulating meeting. The work of C.M. is funded by a Ci\^encia2007 Research Contract, supported by FSE and POPH-QREN funds.
\end{acknowledgements}

\bibliographystyle{aa}
\bibliography{jd9_nunes}

\end{document}